\documentclass[trackchanges,twocolumn]{aastex701}

\usepackage[utf8]{inputenc}
\usepackage[T1]{fontenc}
\usepackage{gensymb}
\usepackage{siunitx}
\usepackage{lineno}
\usepackage{amsmath}
\usepackage{enumitem}

\graphicspath{{./}{figures/}}

\begin{document}

\title{Dynamically driven collapse of a thermally stable accretion disk in a nova-like system}

\author[orcid=0009-0006-2521-033X]{Wentao Li}
\affiliation{Department of Astronomy, University of Science and Technology of China, Hefei 230026, People's Republic of China}
\affiliation{School of Astronomy and Space Sciences, University of Science and Technology of China, Hefei 230026, People's Republic of China}
\email{wtli@mail.ustc.edu.cn}

\author[orcid=0000-0003-3965-6931]{Jie Lin}
\affiliation{Department of Astronomy, University of Science and Technology of China, Hefei 230026, People's Republic of China}
\affiliation{School of Astronomy and Space Sciences, University of Science and Technology of China, Hefei 230026, People's Republic of China}
\email{linjie2019@ustc.edu.cn}

\author[orcid=0000-0002-1517-6792]{Tinggui Wang}
\affiliation{Department of Astronomy, University of Science and Technology of China, Hefei 230026, People's Republic of China}
\affiliation{School of Astronomy and Space Sciences, University of Science and Technology of China, Hefei 230026, People's Republic of China}
\affiliation{College of Physics, Guizhou University, Guiyang 550025, People's Republic of China}
\email{twang@ustc.edu.cn}

\author[orcid=0000-0002-3935-2666]{Yongkang Sun}
\affiliation{National Astronomical Observatories, Chinese Academy of Sciences, Beijing 100101, People’s Republic of China}
\affiliation{School of Astronomy and  Space Science, University of Chinese Academy of Sciences, Beijing 100049, People’s Republic of China }
\email{sunyk@bao.ac.cn}

\author[orcid=0000-0002-2452-551X]{Chengyuan Wu}
\affiliation{Yunnan Observatories, Chinese Academy of Sciences, Kunming 650216, People’s Republic of China}
\affiliation{Key Laboratory for Structure and Evolution of Celestial Objects, Chinese Academy of Sciences, Kunming 650216,
People’s Republic of China}
\affiliation{International Centre of Supernovae, Yunnan Key Laboratory, Kunming 650216, People’s Republic of China}
\email{wuchengyuan@ynao.ac.cn}

\author{Heran Xiong}
\affiliation{Research School of Astronomy and Astrophysics, Australian National University, Canberra, ACT2611, Australia}
\email{xiongheran@gmail.com}

\author[orcid=0000-0002-4005-5095]{Krystian I\l{}kiewicz}
\affiliation{Nicolaus Copernicus Astronomical Center, Polish Academy of Sciences, Bartycka 18, 00-716 Warsaw, Poland}
\email{krystian.ilkiewicz@gmail.com}

\author[orcid=0000-0003-2688-7511]{Luca Casagrande}
\affiliation{Research School of Astronomy and Astrophysics, Australian National University, Canberra, ACT2611, Australia}
\affiliation{ARC Centre of Excellence for All Sky Astrophysics in 3 Dimensions (ASTRO 3D), Australia}
\email{luca.casagrande@anu.edu.au}

\author{Michael Bessell}
\affiliation{Research School of Astronomy and Astrophysics, Australian National University, Canberra, ACT2611, Australia}
\affiliation{ARC Centre of Excellence for All Sky Astrophysics in 3 Dimensions (ASTRO 3D), Australia}
\email{michael.bessell@anu.edu.au}

\author[orcid=0000-0002-7152-3621]{Ning Jiang}
\affiliation{Department of Astronomy, University of Science and Technology of China, Hefei 230026, People's Republic of China}
\affiliation{School of Astronomy and Space Sciences, University of Science and Technology of China, Hefei 230026, People's Republic of China}
\email{jnac@ustc.edu.cn}

\author[orcid=0000-0003-4700-348X]{Chengyi Wang}
\affiliation{Department of Astronomy, University of Science and Technology of China, Hefei 230026, People's Republic of China}
\affiliation{School of Astronomy and Space Sciences, University of Science and Technology of China, Hefei 230026, People's Republic of China}
\email{chengyi_wang@ustc.edu.cn}

\correspondingauthor{Jie Lin, and Tinggui Wang}
\email{linjie2019@ustc.edu.cn; twang@ustc.edu.cn}

\begin{abstract}

While standard disk instability theory predicts thermal-steady, outburst-free disks in nova-like variables due to their high mass transfer rates, mass transfer variations are prevailingly invoked to explain the occasional flaring/fading phenomena observed in these systems.
Here we report the observational evidence for a dynamical collapse of a thermally stable accretion disk, captured serendipitously by TESS during a fading episode of the VY Scl-type nova-like system MASTER~OT~J072703.91-631952.8, and traced by the emergence of an unusual negative superhump that evolves toward its orbital frequency.
Simultaneously, the system underwent an anomalous eruptive event featuring a remarkably symmetric 45-day light-curve profile, indicative of a mild energy-release mechanism fundamentally distinct from documented eruptive events in cataclysmic variables.
Notably, the concurrence of the eruption and the disk collapse is difficult to reconcile with the paradigm of mass transfer variations: the eruption implies enhanced mass transfer, whereas the disk collapse indicates a decline. 
The morphological disk evolution over $\sim 700$~days, characterized by both negative and positive superhumps, indicates a transition from a circular disk to an eccentric disk, followed by a tilted state and ultimately a minimal disk configuration.
This evolutionary sequence provides evidence for a previously unrecognized dynamics-driven cycle operating in the thermally stable accretion disk of a VY Scl-type nova-like star.
\end{abstract}

\section{Introduction} \label{sec:intro} 

VY~Scl variables are a unique subclass of nova-like stars (NL), characterized by irregular, dramatic brightness declines of several magnitudes persisting for weeks to hundreds of days \citep{Warner+1995+CV_book,Leach+etal+1999+VYScl,Honeycutt+Kafka+2004+VYScl}. 
Their accretion disks are thought to be at least partly dissipated in the intermediate or low states \citep{Hameury+etal+2002+VY_model, Hameury+Lasota+2005+VY_radius_tidal}. 
This scenario is supported by observational evidence:
eclipse mapping showed that DW UMa had a reduced disk radius in the intermediate state \citep{Stanishev+etal+2004+eclipse_mapping_DWUMa}; 
combined UV/optical spectral modeling indicated that TT Ari possessed a largely absent accretion disk during its low state \citep{Gansicke+etal+1999+TTAri_low_state};
and Doppler tomography of SDSS J154453.60+255348.8 revealed the disk to be completely missing during its low state \citep{Medina_Rodriguez+eatl+2023+VY_tomography}.

The prevailing interpretation attributes the fading episodes to an abrupt reduction in the mass transfer rate from the donor star, induced either by starspots drifting underneath the inner Lagrangian point \citep[$L_1$,][]{Livio+Pringle+1994+starspots} or a thickening outer disk rim disrupting irradiation feedback mechanisms \citep{Wu+etal+1995+feedback}.
However, since current observations do not provide compelling support for this interpretation, whether changes in the mass transfer rate are indeed the primary driver of the low states remains an open question, and alternative physical mechanisms cannot yet be excluded.

Superhumps are periodic photometric modulations commonly observed in cataclysmic variables (CVs). They are classified as positive superhump (pSH) or negative superhump (nSH), depending on whether their periods are longer or shorter than the orbital period. 
Positive superhumps are thought to arise from the prograde precession of an eccentric accretion disk,  which develops through tidal interactions with the orbital secondary once the disk expands to the 3:1 Lindblad resonance radius \citep{Whitehurst+1988+superhump,Lubow+1991+simulation_superhump,Lubow+1991+model_eccentric}. 
This mechanism is expected to operate only in systems with sufficiently low binary mass ratios.
\cite{Whitehurst+King+superhumps_massratio_limit} estimated a critical value of $q_c=0.33$, \cite{Pearson+2006+resonance_massratio} proposed another $q_c=0.39$, while \cite{smak+2020+resonance_massratio} suggested a substantially lower value of $q_c=0.22$. 
Nevertheless, several systems exhibiting pSHs have been found to possess mass ratios far above the theoretical upper limit \citep{Gies+etal+2013+KIC9406652,Bruch+2023+superhump_NLs}.

In contrast, negative superhumps are generally attributed to the retrograde nodal precession of a tilted accretion disk \citep{Bonnet+Motch+Mouchet+1985+TVCol_nsh,Harvey+etal+1995+V503_nSH}.
This interpretation has been supported by smoothed particle hydrodynamics (SPH) simulations \citep{Wood+etal+2009+SPH_superhump, Montgomery+2009+SPH} and further developed through analytical studies  \citep{ Larwood+1998+nSH_model,Montgomery+2009+earth_moon_nSH,Osaki+Kato+2013+study_DIM}.
Although the tilted-disk interpretation of nSHs is widely accepted, the physical mechanism responsible for producing the disk tilt remains unclear \citep{Montgomery+etal+2010+cause_nSH}.
Several mechanisms have been proposed to produce disk tilt (or warping), including magnetic fields from the secondary star \citep{Barrett+1988+sweeping_nsh}, irradiation by the primary \citep{Pringle+1996+irradiation_disk}, misalignment between the spin and disk axes \citep{Kumar+1986+spin_twist,Roberts+1974+HerX1_spin}, and tidal interactions in the binary system \citep{lubow+Pringle+1993+wave_tilt,Larwood+Papaloizou+1997+SPH_tilt}. 
However, both theoretical considerations and observations suggest that disk tilt may be intrinsically linked to disk eccentricity.
Theoretically, \cite{lubow+1992+eccentric_tilt} demonstrated that, the coupling between horizontal and vertical oscillations at the Lindblad resonances allows disk inclination instability to grow together with the tidal eccentric instability.
More recently, a comprehensive study of the precession behavior in TT Ari further indicates that nSHs tend to emerge when pSHs reach their maximum period excess, implying that disk tilt may develop as the disk expands to a critical radius \citep{Suleimanov+etal+2024+nSH_TTAri}.

In this paper, we present a comprehensive view of the morphological evolution of the accretion disk throughout a brightness drop in a VY Scl variable, MASTER~OT~J072703.91-631952.8 \citep[$\alpha=111.76617$, $\delta=-63.33125$, hereafter J0727,][]{ATel+8190+2015,ATel+8191+2015}, aiming at investigating the physical mechanism responsible for the onset of the low state.
Benefit from intensive coverage  by \textit{TESS}, we resolve pSHs and nSHs in the photometry, which characterize the eccentric and tilted morphologies of the accretion disk, respectively.
The photometric and spectroscopic analyses, WWZ and Lomb-Scargle periodogram computation, and the binary solution are presented in Section~\ref{sec:data analysis}.
In Section~\ref{sec:discussion} we track the disk collapse, locate the accompanying eruption in duration-energy diagram, and propose a dynamics-driven morphological cycle to interpret the collapse of thermally stable accretion disk. A final summary is given in Section~\ref{sec:conclusion}.

\section{Data Analysis \& Result} \label{sec:data analysis}

\subsection{Photometric and spectroscopic analysis} \label{subsec:photometry and spectroscopy}
\begin{figure*}
\centering
    \includegraphics[width=0.95\textwidth]{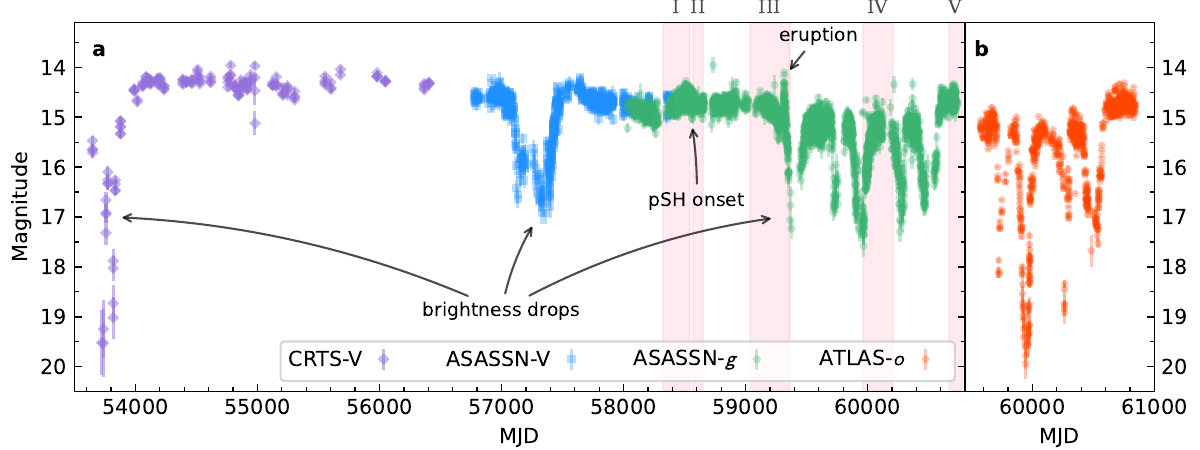}
    \caption{Long-term optical monitoring of J0727. 
    a, Decades light curves of J0727 consisted of CRTS (V band) and 
    ASASSN (V and $g$ bands). 
    Red shaded regions denote five series of TESS observation. 
    b, ATLAS-$c$ band light curves of J0727. Different depths of the brightness drops between ASASSN-$g$ and ATLAS-$c$ bands are attributed to the contamination from a nearby G dwarf ($13\arcsec$ away), which ASASSN cannot resolve from the target.
    The photometric data are presented as mean values $\pm 1\sigma$.
     } 
    \label{fig:archival_lightcurve}
\end{figure*}
The long-term photometric monitoring of J0727, spanning nearly two decades with weekly/monthly cadence, incorporates data from All-Sky Automated Survey for Supernovae \citep[{ASASSN},][]{Shappee+etal+2014+ASASSN,Kochanek+etal+2017+ASASSN}, Catalina Real-Time Transient Survey \citep[{CRTS},][]{Drake+etal+2009+CRTS} and Asteroid Terrestrial-impact Last Alert System \citep[{ATLAS},][see Appendix~\ref{subsec:long-term photometry}]{Tonry+etal+2018+ATLAS,Heinze+etal+2018+ATLAS+variables,Smith+etal+2020+ATLAS}. 
As shown in Fig.~\ref{fig:archival_lightcurve}, J0727 exhibits characteristic 2--5 mag irregular fading in absence of a stable low accretion state, matching fundamental light-curve features of VY Scl variables. 
The dereddened colour and absolute magnitude of J0727 derived from {\it Gaia} DR3 dataset \citep{Gaia_Collaboration+2016+performance,Gaia+DR3+2022,GaiaDR3+Apsis+parameters} are $(B_{\rm p}-R_{\rm p})_0=0.020\pm0.094$\,mag and $M_{\rm G}=4.410\pm0.170 $\,mag, respectively (see Appendix~\ref{subsec:Gaia CMD}), which align with nova-like variables in {\it Gaia} colour-magnitude diagram \citep{Abril+etal+2020+Gaia_CV_distribution}.
Optical spectrum yields a single-component spectrum (see Appendix~\ref{sec:spectroscopy}) dominated by hydrogen and helium emission lines, confirming J0727 as an accreting binary. The prominent He~II $\lambda$4686 line further indicates a high effective temperature of J0727, agreeing with the characteristic of VY~Scl stars \citep{Weil+etal+2018+VY_two_spectra,Medina_Rodriguez+eatl+2023+VY_tomography}.

Besides, as a high-ecliptic-latitude source, J0727 has been persistently monitored for nearly a year at 2--30 minute cadence by Transiting Exoplanet Survey Satellite \citep[TESS,][]{Ricker+etal+2014+TESS,TESS+2015}. We describe the reduction of TESS data in Appendix~\ref{subsec:TESS photometry}.
The high-cadence observations coincidentally encompassed the third brightness drop (hereafter BD2021, series~III in Fig.~\ref{fig:archival_lightcurve}), allowing us to resolve the detailed temporal evolution during the 45-day eruptive event (around MJD~59320, hereafter E202104).

\subsection{WWZ analysis and Lomb-Scargle periodogram} \label{subsec:WWZ and LSP}
\begin{figure*}
\centering
    \includegraphics[width=0.95\textwidth]{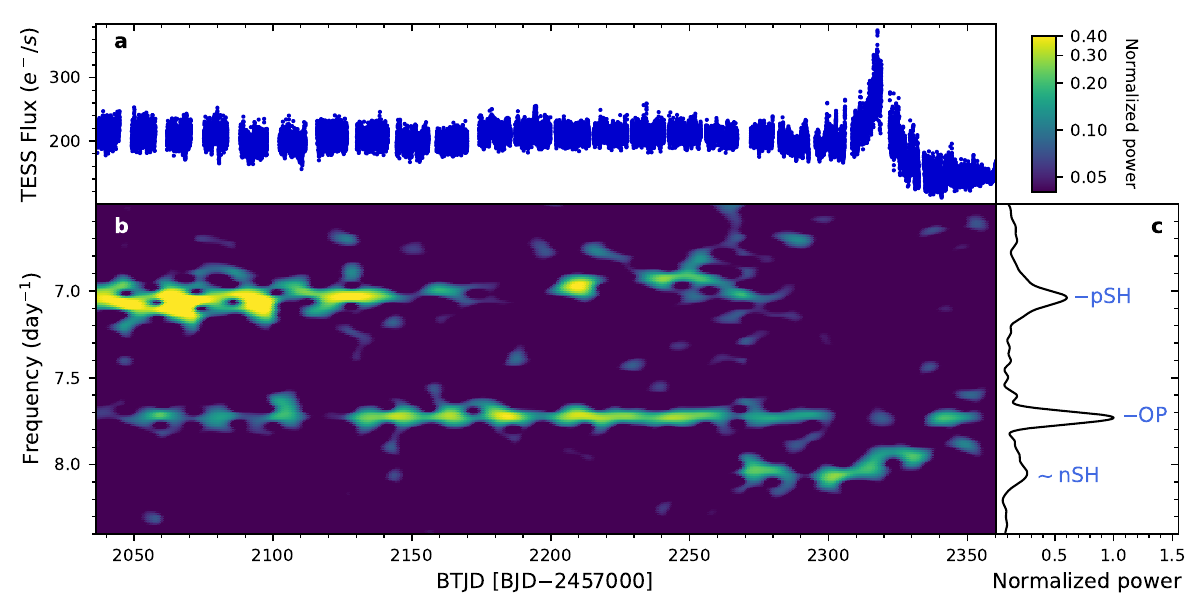}
    \caption{Normalized WWZ and stacked periodogram covering the brightness drop BD2021. 
    a, 
    TESS PDCSAP light curve at 600-second cadence obtained from {\it TESS-SPOC}. 
    b, Normalized WWZ plot computed from the TESS light curve.
    The WWZ powers were normalized by its global maximum value.
    The frequency bin width is $0.01~{\rm day}^{-1}$, and the time bin width is 1.0 day. 
    The WWZ peaks in each time bin indicate the characteristic frequencies for that epoch.
    c, Stacked Lomb-Scargle periodogram (LSP).
    The periodogram was produced by coadding LSPs from 24 segments (see Appendix~\ref{sec:computation of WWZ and LSP}) partitioned by observational gaps, and then normalized by its maximum power.
    Three principle frequency features are labeled.
    } 
    \label{fig:wwz}
\end{figure*}
To visualize the time-dependent, short-timescale variability from the TESS observation data, we computed weighted wavelet Z-transform \citep[{WWZ},][]{Foster+1996+WWZ} using the {\tt wwz} Python package \citep{Kiehlmann+Max-Moerbeck+King+2023+wwz_package}, as shown in Fig.~\ref{fig:wwz}. Given that the characteristic frequencies presented from the WWZ analysis are inherently parameter-dependent and mutable, we thus determined precise feature frequencies by stacking Lomb-Scargle periodograms \citep[LSPs,][]{Lomb+1976+lomb_scargle,Scargle+1982} computed for each individual TESS observational segment.
The computation of WWZ and Lomb-Scargle periodogram is introduced in Appendix~\ref{sec:computation of WWZ and LSP}.

The WWZ analysis identified three principal periodic components:
positive superhump at $\nu_{\rm pSH}=7.048\pm0.005~{\rm day^{-1}}$, orbital period at $\nu_{\rm orb}=7.726\pm 0.004~{\rm day^{-1}}$, and a highly variable negative superhump $\nu_{\rm nSH}\sim8.0~{\rm day^{-1}}$.
The period excess of pSH, $\varepsilon_{+}=(P_{\rm pSH}-P_{\rm orb})/P_{\rm orb}=0.0962\pm0.0009$, 
conforms to the $P_{\rm orb}$-$\varepsilon_{+}$ correlation given by smoothed particle hydrodynamics (SPH) simulations \citep{Wood+etal+2009+SPH_superhump}, validating the identifications for the periodic features.
The time series of nSH frequencies from the LSPs  were converted into frequency excesses $\varepsilon^*_-$ with $\varepsilon^*_- = (\nu_{\rm orb}-\nu_{\rm nSH})/\nu_{\rm orb} $~.

\subsection{Binary solution} \label{subsec:binary solution}

As the key parameter for determining the natures of CV systems, mass ratio $q=M_2/M_1$ can be estimated through the $\varepsilon$-$q$ relations in three independent approaches as follows:
\begin{itemize}
    \item  Precessing-disk model. 
    Following the established theoretical framework that pSH arise from a precessing eccentric disk at the 3:1 resonance radius \citep{Whitehurst+1988+superhump}, the pSHs observed in superoutburst of DN can be classified into three stages and used to estimate mass ratio for stage A and stage B \citep{Kato+etal+2009+SH_evolution_in_DN,Kato+etal+2013+precession}. Assuming that the pSH prior to E202104 corresponds to the stage A, the mass ratio can then be estimated following the formula,
    \begin{equation*}
        q = -0.0016 + 2.60\varepsilon_+^*+3.33(\varepsilon_+^*)^2+79.0(\varepsilon_+^*)^3~,
    \end{equation*}
    where $\varepsilon_+^*=\nu_{\rm aPR}/\nu_{\rm orb}=1-\nu_{\rm pSH}/\nu_{\rm orb}$ is the frequency deficit of pSH.
    We obtained $q = 0.306\pm0.004$ for J0727.
    \item Smoothed partical hydrodynamics (SPH) simulation.
    From the numerical SPH simulation of accretion disk dynamics \citep{Wood+etal+2009+SPH_superhump},  the relationship between pSH period excess ($\varepsilon_+$) and mass ratio is modeled as
    $q=3.733\varepsilon_+-7.898\varepsilon_+^2~$.
    Applying this to the feature frequencies of J0727 yielded $q=0.284\pm0.002$. 
    \item Empirical relation. By solving the empirical relation for pSH period excess and mass ratio in both DN and NLs \citep{Patterson+etal+2005+precession}, $\varepsilon_+ = 0.18q+0.29q^2$, we obtained an empirical mass ratio of $q = 0.344\pm0.002$ for J0727.
\end{itemize}
It should be noted that the uncertainties presented here account solely for the error originating from the characteristic frequencies.
Given that the derived mass ratio show significant dispersion among methods, we adopt a representative value of $q=0.306\pm0.030$, which is broadly compatible with all three solutions.

As a Roche-lobe-filling component, the donor star in J0727 has a radius ($R_{2}$) approximately equal to its Roche lobe radius ($R_{L,2}$), such that $R_{2} \approx R_{L,2}$.
With the Eggleton's approximation \citep{Eggleton+1983+roche_lobe}, the Roche lobe radius follows
\begin{equation}
    r_{L,2} \equiv \frac{R_{\rm L,2}}{a} \approx \frac{0.49q^{2/3}}{0.6q^{2/3}+\ln(1+q^{1/3})}~,
    \label{Eq:roche lobe radius}
\end{equation}
where $r_{L,2}$ denotes the dimensionless Roche lobe radius, normalized to the orbital separation $a$.
Hence, the donor star's radius is $R_2 \approx R_{\rm L,2} \approx 0.282\, a$.
Following the semi-empirical mass–radius relation for CV donor stars \citep{Knigge+Baraffe+Patterson+2011+CV_donor_sequence}, the donor radius together with orbital period yields a donor mass of $M_2 \approx 0.20\,{\rm M_{\odot}}$. Combined with the estimated mass ratio, this implies a primary mass $M_1 \approx 0.65\,{\rm M_{\odot}}$. Kepler's third law then gives an orbital separation $a \approx 7\times10^{10}\,\rm cm$. Since these binary parameters are ultimately anchored from the adopted mass ratio, their uncertainties are governed primarily by the uncertainty in $q$.

\section{Discussion} \label{sec:discussion}

\subsection{Tracking the collapse of accretion disk} \label{subsec:floats}

\begin{figure}
\centering
    \includegraphics[width=0.45\textwidth]{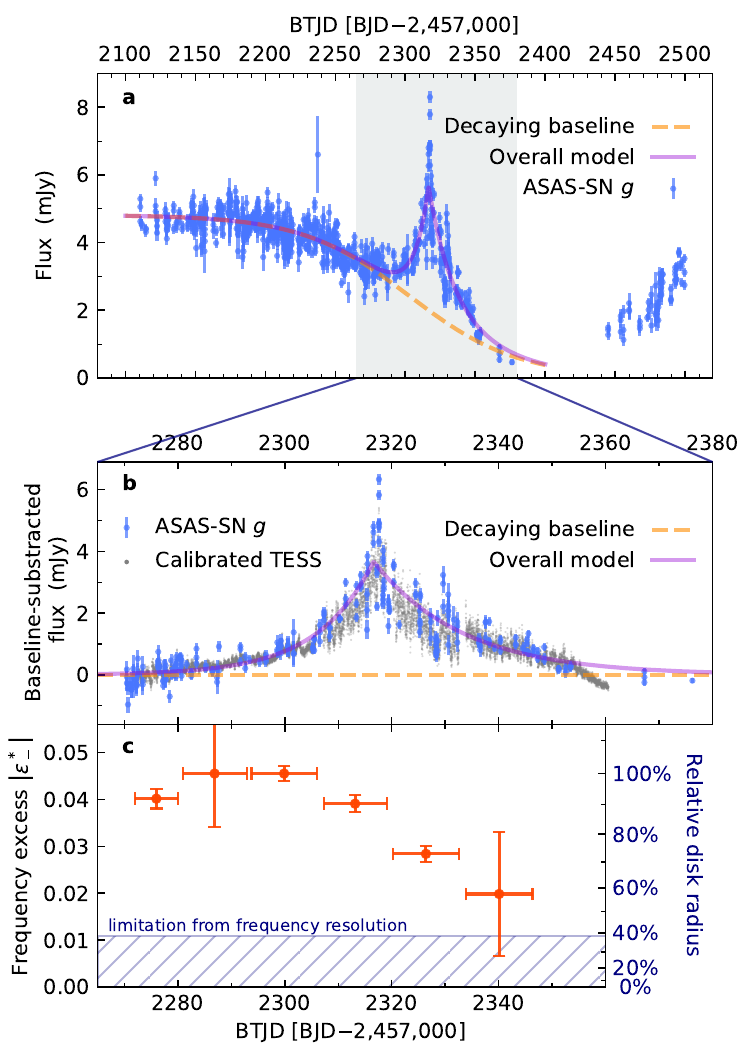}
    \caption{Light curve and frequency evolution of negative superhump (nSH) during eruption E202104.
    a, ASASSN-$g$ light curve (blue markers) covering both the brightness drop BD2021 and eruption E202104. 
    The best-fitting model (purple solid line) consists of a logistic function (for the decaying baseline; the orange dotted line; Appendix~\ref{sec:lightcurve fitting}) and a double-exponential profile (for the eruption).
    b, Zoom-in view of baseline-subtracted ASASSN-$g$ light curve. 
    The TESS SAP light curves (grey points), calibrated to ASASSN-$g$ fluxes (see Appendix~\ref{sec:eruption statistics}), are overplotted for comparison.
    c, Evolution of nSH frequency excess ($\varepsilon^*_-$; red markers) synchronized with eruption E202104.
    The $\varepsilon^*_-$ values were derived from Lomb-Scargle periodograms computed over segments spanning 7.9 day to 12.5 day.
    The additional right axis denotes the corresponding relative disk radius derived from $\varepsilon^*_-$, where 100\% represents the maximum disk radius.
    The shaded region indicates the limit imposed by frequency resolution of Lomb-Scargle periodogram (see Appendix~\ref{sec:computation of WWZ and LSP}). Horizontal error bars represent the time intervals of each segment analyzed; vertical error bars represent the 1$\sigma$ confidence intervals.
     } 
    \label{fig:eruption_radius_decaying}
\end{figure}

Negative superhumps are commonly employed to constrain accretion disk sizes, as a widely accepted methodology applied across X-ray binaries \citep{Larwood+1998+nSH_model,Brocksopp+etal+1999+CygX1+nSH,Wijers+Pringle+1999+nSH_Warped_disk} and CVs \citep{Osaki+Kato+2013+cause_DIM,Osaki+Kato+2013+study_DIM,Kato+Hiroyuki+2013+nSH_KIC8751494,Ohshima+etal+2014+ERUMa_nSH,Suleimanov+etal+2024+nSH_TTAri}.
The nodal precession rate $\nu_{\rm nPR}$ of a tilted disk relative to the orbital frequency \citep{Osaki+Kato+2013+study_DIM} follows
\begin{equation}
\varepsilon^*_- \equiv \frac{\nu_{\rm nPR}}{\nu_{\rm orb}} =-\eta \frac{3}{7} \frac{q}{\sqrt{1+q}} \left (\frac{R_{\rm d}}{a} \right )^{3/2} \times \cos{\theta}~,
\label{Eq:precession_rate}
\end{equation}
where $R_{\rm d}$ denotes the disk radius, $\theta$ represents the tilt angle (typically $ \cos{\theta} \approx 1$), and $\eta$ is a correction factor encoding different mass distribution within the disk \citep[e.g. $\eta=1.09$ for a homogeneous disk,][]{Osaki+Kato+2013+study_DIM,Montgomery+2009+earth_moon_nSH}.
Due to the unknown mass distribution preventing the unique determination of $\eta$ in this study,
the Eq.~\ref{Eq:precession_rate} is simplified to a power-law dependence \citep{Kato+Hiroyuki+2013+nSH_KIC8751494}, 
 $\lvert \varepsilon^*_- \rvert \propto R_{\rm d}^{3/2}$.

As illustrated in Fig.~\ref{fig:eruption_radius_decaying}, the frequency excess of nSHs ($\lvert \varepsilon^*_- \rvert$) exhibited a sharp decrease synchronized with the eruptive event, E202104.
Over 45 days, the $\lvert \varepsilon^*_- \rvert$ declined from $\approx 0.046$ to $\approx 0.020$, indicating rapid disk contraction to <60\% of its pre-eruption radius.
While the disk size could potentially shrink further, the finite frequency resolution in the Lomb-Scargle periodograms precluded subsequent measurements (see Appendix~\ref{sec:computation of WWZ and LSP}).
Given that the accretion disk dominates high-state luminosity, the disk collapse naturally triggered the characteristic low state observed in the light curve.
Since low states in VY Scl variables are attributed to the absence \citep{Medina_Rodriguez+eatl+2023+VY_tomography} or small scale \citep{Schmidtobreick+etal+2018+small_disk} of the accretion disk, 
such disk collapse constitutes an inherent process bridging the full-scale disk and the minimal disk state.

\subsection{Prolonged eruptive event during the disk collapse} \label{subsec:tables}

\begin{figure}
\centering
    \includegraphics[width=0.45\textwidth]{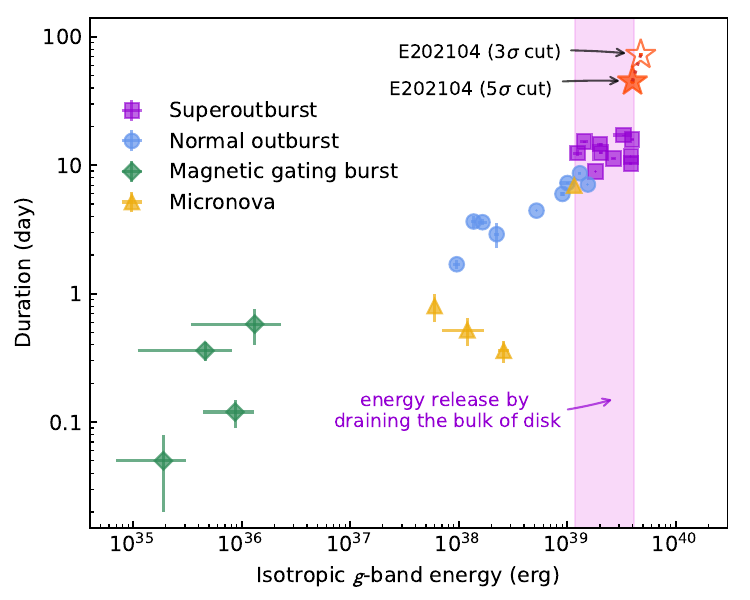}
    \caption{Duration-energy distribution for various types of eruptive events. 
    The eruption samples here include normal outburst \citep[blue circles,][]{Ilkiewicz+etal+2024+DN_outburst}, superoutburst (purple squares), magnetic gating burst \cite[green diamonds,][]{Scaringi+etal+2017+Nature+type2,Littlefield+etal+2022+type2,Scaringi+etal+2022+switch_type2,Ilkiewicz+etal+2024+DN_outburst}, micronova \cite[yellow triangles,][]{Scaringi+etal+2022+Nature+type1,Veresvarska+etal+2024+bursting_type1}, and eruption E202104 (red star).
    All eruptive events were measured using a 5$\sigma$ threshold cut above the quiescence (see Appendix~\ref{sec:eruption statistics}).
    Additionally, E202104's marker under 3$\sigma$ threshold cut is also shown, since its duration would be significantly underestimated using the 5$\sigma$ cut due to its mild eruptive profile.
    The purple shaded region indicates the typical energy range of superoutbursts, representing the energy released by draining the bulk of the accretion disk \citep{Osaki+1996+TTI,Osaki+Kato+2013+cause_DIM}.
    The durations and energies are presented as mean values $\pm 1\sigma$.
     } 
    \label{fig:duration_energy_distribution}
\end{figure}

As shown in Fig.~\ref{fig:eruption_radius_decaying}, eruptive event E202104 emerged during the decaying phase between high and low states.
Its excess flux above the decaying baseline is well modeled by a double-exponential profile, 
yielding a duration of 45~days and an excess energy of $4\times 10^{39}$~erg (see Appendix~\ref{sec:lightcurve fitting}).
As shown in Fig.~\ref{fig:duration_energy_distribution}, E202104 fundamentally differs from previously documented outbursts/superoutbursts \citep{Otulakowska-Hypka+etal+2016+DN_statistic,Ilkiewicz+etal+2024+DN_outburst}, 
localized thermonuclear bursts \citep{Scaringi+etal+2022+Nature+type1,Veresvarska+etal+2024+bursting_type1} as well as magnetically gated accretion \citep{Scaringi+etal+2017+Nature+type2,Littlefield+etal+2022+type2,Scaringi+etal+2022+switch_type2,Ilkiewicz+etal+2024+DN_outburst} .
The three-week rise phase implies a mild energy release mechanism, incompatible with thermonuclear burning runaway \citep{Scaringi+etal+2022+Nature+type1} or disk thermal instability \citep{Lasota+2001+NewAR+DIM,Hameury+2020+review}.
Its total released energy is comparable to those of superoutbursts, which nearly exhaust the accretion disk \citep{Osaki+1996+TTI,Osaki+Kato+2013+cause_DIM}.
Although the symmetric temporal profile resembles milinovae \citep{Maccarone+etal+2019+SSS,Hillman+etal+2019+milinovae,Kato+etal+2020+ASASSN16oh,Mroz+etal+2024+milinova,Hachisu+Kato+2025+milinova}, 
its peak luminosity ($4\times 10^{33}$~erg/s) is approximately 1000 times fainter, indicating an accretion rate insufficient for steady hydrogen burning \citep{Kato+etal+2020+ASASSN16oh}.
Furthermore, we have identified similar eruptions at the onset of the high-to-low state transition in several other VY Scl variables (e.g. MGAB-V239 and V425 Cas), suggesting that E202104-like events may be more common than previously recognized (Li, Lin et al., in preparation).
The non-unique occurrence strongly disfavors a chance microlensing origin \citep{Sajadian+Jorgensen+2022+microlensing,Mao+2012+review_microlensing}.

Beyond the physical origins of eruptive events discussed above, mass transfer variations are widely invoked to explain both stunted outbursts \citep{Honeycutt+etal+1998+Unusual_Stunted,Robertson+etal+2018+Stunted_MT,Duffy+etal+2024+VY_detail,Honeycutt+etal+2025+stunted_outbursts} and month-long flares \citep{Zsidi+etal+2023+trio} in NLs, as well as the characteristic brightness drops of VY Scl variables \citep{Wu+etal+1995+feedback,Leach+etal+1999+VYScl,Hameury+etal+2002+VY_model,Hameury+Lasota+2005+VY_radius_tidal}.
Analogous to IW And-type behavior in anomalous Z~Cam stars \citep{Hameury+Lasota+2014+IW_and}, the eruptive event E202104 here requires an episode of enhanced mass transfer, whereas the following brightness drops are commonly attributed to a declining mass transfer rate.
However, in this case, the collapse of accretion disk, which represents an intrinsic process leading to the brightness drop, occurred synchronously with the eruption.
This simultaneity raises a fundamental paradox: the eruption necessitates enhanced mass transfer, but the disk collapse requires a declining rate.

Although the thermal instability model predicts that outbursts would be triggered under a slowly decaying mass transfer \citep{Hameury+etal+2002+VY_model}, these predicted day-scale events fundamentally differ from the observed 45-day eruptive event.
Overall, E202104 cannot be simply explained by mass transfer variations, disk thermal instability, or thermonuclear burning.
Given its temporal coincidence with the disk collapse and an energy release comparable to superoutbursts, we speculate that the eruption is most likely powered by the gravitational energy liberated during the dramatic contraction of the accretion disk. However, it remains unclear why similar events have so far been observed only in a small fraction of VY Scl stars.
To account for the physical mechanism driving the disk collapse, we propose that, alongside the well-established thermal-viscous limit cycle \citep{Smak+1984+DIM, Lasota+2001+NewAR+DIM}, 
a previously unrecognized dynamics-driven cycle operates in accreting compact binaries, characterized by the morphological evolution of the accretion disk.

\subsection{A dynamics-driven morphological cycle of the accretion disk} 
\label{subsec:TTI in TSD}

\begin{figure*}
\centering
    \includegraphics[width=0.95\textwidth]{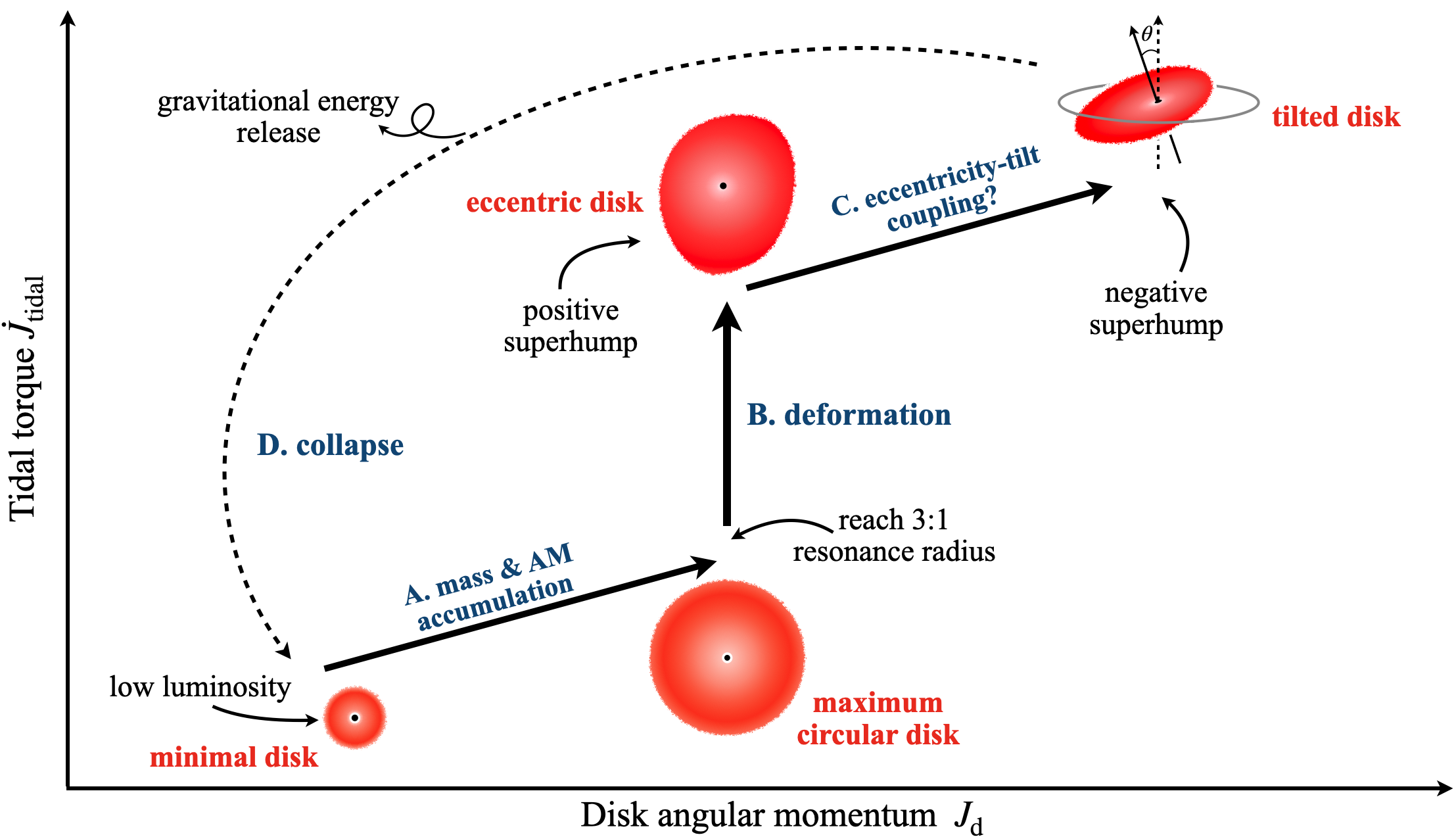}
    \caption{Schema of dynamics-driven morphological cycle in a thermally stable accretion disk. 
    Disk morphologies are depicted using perfect (circular disk) , distorted (eccentric disk) and inclined (tilted disk) circles, with sizes scaled schematically.
    Straight thick arrows illustrate the evolutionary stages of the dynamics-driven cycle:
    A. mass and AM transfer from the donor star increases the disk size from the minimal state through the maximum stable circular configuration;
    B. upon reaching the 3:1 resonance radius, the maximum circular disk deforms into an eccentric disk;
    C. the eccentric disk evolves into a tilted disk through a yet unclear mechanism (e.g. eccentricity–tilt coupling);
    D. tilted accretion disk catastrophically collapses back to the minimal circular state.
    The black dashed arrow denotes the evolution stage corresponding to the disk collapse (BD2021), associated with the eruption E202104.
     } 
    \label{fig:tilt-disruption_cycle_scheme}
\end{figure*}

By solving the thermal equilibrium equations of accretion disk annuli, the standard disk instability model establishes an ``S-shaped'' thermal equilibrium curve between the surface density and mass accretion rate \citep{Smak+1984+DIM,Hameury+etal+1998+DIM_new_version,Lasota+2001+NewAR+DIM,Hameury+2020+review,Kolb+book+2010}. 
The curve comprises three regimes: two stable hot/cool branches flanking an unstable intermediate branch.
The normal outbursts in DNe are interpreted as thermal limit-cycle oscillations between two stable branches when their mass transfer rates reside between the stable branches (i.e., $\dot{M}_{\rm cr, cool} <\dot{M}_{\rm T}<\dot{M}_{\rm cr, hot}$).

Although an analytical formulation of tidal dissipation remains elusive \citep{Papaloizou+Pringle+1977+tidal_torque,Ichikawa+Osaki+1994+tidal_torque}, 
the thermal-tidal instability (TTI) model imaginatively proposes an analogous S-shaped angular momentum (AM) equilibrium curve in the $J_{\rm d}$-$\dot{J}_{\rm tidal}$ parameter space \citep{Osaki+1996+TTI,Osaki+2005+DIM+review}, 
where $J_{\rm d}$ denotes total AM of the accretion disk and $\dot{J}_{\rm tidal}$ is angular momentum loss (AML) rate due to tidal torques.
This curve features three phases: a stable circular-disk phase with weak tidal torques, a stable eccentric-disk phase governed by amplified tidal torques, and an unstable transitional phase. 
Theoretically, the superoutbursts in DNe arise from tidal limit-cycle oscillations between the circular- and eccentric-disk phases, 
when normal outbursts expand the disk  to 3:1 Lindblad resonance radius \citep{Whitehurst+1988+superhump, Lubow+1991+model_eccentric,Lubow+1991+simulation_superhump} via enhanced viscosity.
As a promising framework to unify various outburst behaviours in dwarf novae (DNe), TTI could potentially be extended to nova-like variables (NLs), while the fundamental dichotomy (between DNe and NLs) stems from whether the mass transfer rate ($\dot{M}_{\rm T}$) surpasses the critical threshold required to maintain a fully ionized accretion disk \citep{Lasota+2001+NewAR+DIM,Hameury+2020+review,Kolb+book+2010}.
Critically, the TTI model predicts a potential scene in NLs: when high mass transfer rate maintains the entire accretion disk on the thermally stable branch, suppressing the conventional thermal limit cycle, the tidal instability still operates and drive the morphological cycle of accretion disks.

Inspired by the morphological cycle envisioned in the TTI model,we propose an analogous evolutionary picture to explain the mechanism driving disk collapse in J0727. Within this framework, the accretion disk evolves through four stages: 
\begin{description}[leftmargin=0pt, labelindent=0pt, listparindent=2em]
    \item[A. Minimal disk $\Rightarrow$ maximum circular disk ] 
    The minimal-disk state corresponds to the low-luminosity state observed during brightness drops (e.g. $\sim$MJD 57130--57400).
    In this phase, the accretion disk is either severely depleted or nearly absent, as supported by Doppler tomography\citep{Medina_Rodriguez+eatl+2023+VY_tomography} and eclipse mapping\citep{Stanishev+etal+2004+eclipse_mapping_DWUMa}.
    As mass and AM are persistently transferred from the secondary and accumulated in the disk, the initially minimal accretion disk gradually rebuilds into a larger circular configuration, sustaining a higher accretion rate that gives rise to the elevated luminosity observed in the high state (e.g. $\sim$MJD 57460--59200).
    \item[B. Maximum circular disk $\Rightarrow$ eccentric disk]
    The state of maximum circular disk corresponds to the phase when the disk just reaches the 3:1 Lindblad resonance radius \citep{Lubow+1991+model_eccentric,Lubow+1991+simulation_superhump}. 
    Beyond this radius, the disk becomes tidally unstable and is deformed from a circular into an eccentric configuration due to the tidal interaction with the secondary star \citep{Whitehurst+1988+superhump, Lubow+1991+model_eccentric, Osaki+Kato+2013+cause_DIM}.    
    \item[C. Eccentric disk $\Rightarrow$ tilted disk] 
    This eccentric-disk state is characterized by the positive superhumps (pSHs), observed during $\sim$MJD 58573--58682 and $\sim$MJD 59036--59270 (see Appendix.~\ref{sec:computation of WWZ and LSP}), implying that this state likely persisted for approximately 700 days.
    The physical mechanism responsible for the transition from an eccentric disk to a tilted disk remains poorly understood. 
    Theoretically, one proposed mechanism is that the 3:1 Lindblad resonance responsible for exciting disk eccentricity also excites a weaker tilt instability, with a predicted growth rate of $\lambda_i \approx 0.02\,\lambda_e$ \citep{lubow+1992+eccentric_tilt}. 
    Alternatively, the asymmetric geometry of an eccentric disk may generate a net lift torque from the gas stream, gradually tilting the disk \citep{Montgomery+etal+2010+cause_nSH}.
    Observationally, systematic studies of the disk precession in TT~Ari support the occurrence of a pSH-to-nSH transition at a boundary radius \citep{Suleimanov+etal+2024+nSH_TTAri}.
    \item[D. Tilted disk $\Rightarrow$ minimal disk]
    The tilted-disk state is supported by the detection of negative superhumps during $\sim$MJD 59270--59355 \citep{Bonnet+Motch+Mouchet+1985+TVCol_nsh, Harvey+etal+1995+V503_nSH, Wood+etal+2009+SPH_superhump}.
    The subsequent transition from a tilted disk to the minimal-disk stage was traced by the evolution of an unusual negative superhump toward the orbital frequency, together with a sharp decline in brightness (see Fig.~\ref{fig:eruption_radius_decaying}). This decline was synchronized with the eruptive event E202104, which was inferred to be powered by the gravitational energy released during the collapse of the full-scale disk.
    However, the physical mechanism responsible for this transition is still unclear.
    One proposed scenario is that, once the disk tilt becomes sufficiently large, differential nodal precession induced by the companion overwhelms the internal torques that maintain disk coherence, causing the outer disk to break into independently precessing rings \citep{Dogan+etal+2015+tearing_up_discs}.
    The resulting interactions between these rings enhance dissipation and drive a substantially higher accretion rate, potentially accounting for the eruptive event observed during the brightness drop.
    Since the high mass accretion rate (and thus higher effective temperature) maintain the disk fully ionized, the thermal-viscous outburst was suppressed.
    Such enhanced accretion ($\dot{M}_{\rm acc} > \dot{M}_{\rm tr}$) may also accelerate the depletion of the accretion disk, ultimately leading to the minimal-disk state.
\end{description}

Taken together, these four stages outline a self-consistent dynamical evolutionary picture for the accretion disk in J0727, providing a physical framework for understanding the collapse of a thermally stable accretion disk in a VY Scl system.
Nevertheless, current evidence cannot conclusively attribute all brightness drops in J0727 to this morphological evolutionary picture, and whether this mechanism extends beyond J0727 to other VY Scl stars remains unclear.
In a forthcoming study, we will present statistical evidence for a significant association between disk tilting and brightness drops, based on a large sample of nova-like stars, and further show that eruptive events preceding brightness drops are not unique to J0727 (Li, Lin et al., in preparation).

\section{Conclusion} \label{sec:conclusion}

In this paper, we revisit the optical anti-transient
MASTER~OT~J072703.91-631952 (i.e. J0727) and confirm its classification as a VY Scl star through photometric and spectroscopic observations.
Fortunately, a serendipitous 300-day TESS observation captures a brightness-drop episode of J0727, enabling us to investigate the evolution of its short-timescale timing features across the state transition.
By following the evolution of the negative superhump frequency, we track the disk radius and discover the dynamical collapse of J0727's thermally stable accretion disk, representing the inherent evolutionary process between the full-sized and minimal disk states in a VY Scl star.

Concurrent with the disk collapse, we detect an unusual eruptive event characterized by a nearly symmetric temporal profile. Its 45-day duration is substantially longer than previously reported eruptive events in cataclysmic variables, while its total energy of $\sim4\times10^{39}\,\rm erg$ is comparable to that released during superoutbursts, implying that a large fraction of the disk mass is consumed. 
The temporal coincidence between the eruption and the disk collapse is difficult to reconcile with the conventional mass-transfer paradigm, in which both phenomena are attributed solely to variations in the mass-transfer rate.

The long-term evolution of the luminosity and superhump behavior reveals a morphological evolutionary cycle of the accretion disk, progressing from a minimal disk to a large circular disk, then to an eccentric disk, a tilted disk, and finally back to the minimal disk.
We propose this morphological cycle as the manifestation of a dynamics-driven evolutionary process, operating in thermally stable accretion disks alongside the classical thermal limit cycle governed by thermal-viscous instability, providing a promising physical framework for understanding the brightness drops observed in VY Scl stars.

Future observations from facilities like Large Synoptic Survey Telescope \citep{LSST+etal+2009+science_arxiv,Ivezic+etal+2019+LSST}, Wide Field Survey Telescope \citep{WFST+etal+2023+science,Lin+etal+2025+WFST} and Einstein Probe \citep{Yuan+etal+2015+EP} will provide deeper, more diverse data to further investigate disk instabilities in accreting compact binaries.

\begin{acknowledgments}
We are grateful to Simone Scaringi for his useful suggestions.
J.L. is supported by the National Key R\&D Program of China (grant No. 2025YFF0511003),
the National Natural Science Foundation of China (NSFC; Grant No.~12622305 and 12403038), the Fundamental Research Funds for the Central Universities (Grant Numbers WK2030000089), and the Cyrus Chun Ying Tang Foundations.
K.I. was supported by the Polish National Science Centre (NCN) grant 2024/55/D/ST9/01713.

We acknowledge the staff of the ANU 2.3~m telescope and the WiFeS instrument at Siding Spring Observatory for their invaluable technical support: Luca Casagrande and Mike Bessell. Spectrum observation was made with the WiFeS spectrograph on the ANU 2.3~m telescope operated by the Research School of Astronomy \& Astrophysics (RSAA) at the Australian National University.

This paper includes data collected by the TESS mission, which are publicly available from the Mikulski Archive for Space Telescopes (MAST). Funding for the TESS mission is provided by NASA's Science Mission directorate.
The CSS survey is funded by the National Aeronautics and Space Administration under Grant No. NNG05GF22G issued through the Science Mission Directorate Near-Earth Objects Observations Program. The CRTS survey is supported by the U.S.~National Science Foundation under grants AST-0909182 and AST-1313422.
This paper uses data from the ASAS-SN project run by the Ohio State University. We thank the ASAS-SN team for making their data publicly available.
This work has made use of data from the Asteroid Terrestrial-impact Last Alert System (ATLAS) project. The Asteroid Terrestrial-impact Last Alert System (ATLAS) project is primarily funded to search for near earth asteroids through NASA grants NN12AR55G, 80NSSC18K0284, and 80NSSC18K1575; byproducts of the NEO search include images and catalogs from the survey area. This work was partially funded by Kepler/K2 grant J1944/80NSSC19K0112 and HST GO-15889, and STFC grants ST/T000198/1 and ST/S006109/1. The ATLAS science products have been made possible through the contributions of the University of Hawaii Institute for Astronomy, the Queen’s University Belfast, the Space Telescope Science Institute, the South African Astronomical Observatory, and The Millennium Institute of Astrophysics (MAS), Chile.
This work has made use of data from the European Space Agency (ESA) mission {\it Gaia} (\url{https://www.cosmos.esa.int/gaia}), processed by the {\it Gaia} Data Processing and Analysis Consortium (DPAC, \url{https://www.cosmos.esa.int/web/gaia/dpac/consortium}). Funding for the DPAC has been provided by national institutions, in particular the institutions participating in the {\it Gaia} Multilateral Agreement.

We thanks Magdalena Otulakowska-Hypka, Arkadiusz Olech2 and Joseph Patterson for constructing the Dwarf Novae catalog (\url{https://users.camk.edu.pl/magdaot/DN.html}), which is supported by Polish National Science Center grant awarded by decision number DEC-2011/03/N/ST9/03289.
This research has made use of the International Variable Star Index (VSX) database, operated at AAVSO, Cambridge, Massachusetts, USA.

This research made use of Lightkurve, a Python package for Kepler and TESS data analysis \citep{Lightkurve_Collaboration+etal+2018+lightkurve}.
This research has made use of the SVO Filter Profile Service "Carlos Rodrigo", funded by MCIN/AEI/10.13039/501100011033/ through grant PID2023-146210NB-I00.
\end{acknowledgments}

\begin{contribution}
J.L. and W.L. drafted the manuscript, 
T.W. and Y.S. reviewed the manuscript in detail.
T.W. identified the target source and initiated the research direction. 
W.L. discovered the peculiar evolution of nSH, and carried out photometric data collection and analysis.
J.L. interpreted the peculiar nSH, proposed the physical interpretation for the brightness drop, and coordinated the project.
K.I. contributed to the measurements of eruption properties.
L.C. and M.B. obtained and reduced the spectroscopic data.
T.W. Y.S., C.Wu, H.X., N.J., and C.Wang  contributed to the scientific discussion.

\end{contribution}

\setcounter{footnote}{0}
\facilities{CRTS(SSS), TESS, ATLAS, ASAS-SN, ANU(SSO), Gaia}

\software{FIGARO\footnote{\url{https://figaro.readthedocs.io}}, lightkurve(v.2.5)\footnote{\url{https://lightkurve.github.io/lightkurve/}}, tglc(v0.6.6)\footnote{\url{https://github.com/TeHanHunter/TESS_Gaia_Light_Curve}}, wwz\footnote{\url{https://github.com/skiehl/wwz}}, pymc(v.5.22.0)\footnote{\url{https://www.pymc.io}}}

\appendix

\section{Photometric observations} \label{sec:photometrics}

\subsection{Long-term photometry} \label{subsec:long-term photometry}
We collected {\it ASASSN} V- and $g$-band data from Sky Patrol V1.0 via website portal \citep{Shappee+etal+2014+ASASSN, Kochanek+etal+2017+ASASSN}. The CRTS light curves were retrieved from CSDR3 Search Service, and the V-band magnitude is converted from unfiltered photometry\citep{Drake+etal+2013+CRTS_calibration}. We obtained $o$-band light curves from the ATLAS forced photometry server, and these photometric data are further refined by excluding the measurements with SNR$<3.0$ \citep{Rest+etal+2023+ATLAS_filters}.

Given poor spatial resolution of ASASSN \citep[$8\arcsec/{\rm pixel}$; FWHM$\approx 15\arcsec$,][]{Kochanek+etal+2017+ASASSN}, it is necessary to identify the potential contribution of any nearby sources to the observed photometric variations of J0727.
Consequently, a faint contaminating source, located 13$\arcsec$ away from J0727, was revealed from the {\it Gaia} observations with a $G$-band magnitude of $G=16.784\pm0.001$.
Its absolute magnitude and color suggest it as a G-type dwarf star.
While the flux contribution from this G-type dwarf is negligible during the high-luminosity state of J0727, it becomes significant during the low-luminosity state, leading to an overestimation of the observed brightness of J0727 during the decline.
In contrast, the higher spatial resolution from ATLAS \citep[1.86$\arcsec/{\rm pixel}$; FWHM$\approx 3.7\arcsec$,][]{Smith+etal+2020+ATLAS} and CRTS ($1\sim2\arcsec/{\rm pixel}$) effectively resolves this G-type dwarf from J0727, minimizing the contribution from the contaminating source.
Consequently, ATLAS and CRTS light curves reveal significantly deeper brightness minima ($> 19$~mag, see Fig.~\ref{fig:archival_lightcurve}).
Critically, while G-type dwarfs are known to produce flares and superflares \citep{Maehara+etal+2012+superflares}, their characteristic timescales are typically under a few hours \citep{Wu+etal+2015+superflare_saturate, Tu+etal+2020+superflare_TESS}, far shorter than the observed duration of the eruptive event E202104.

\subsection{TESS photometry} \label{subsec:TESS photometry}

To characterize the short-timescale variability of J0727 and further investigate the evolution of its frequency features, we utilized TESS observation data from multiple pipelines, spanning TESS Sector 1 to 90 \citep{Ricker+etal+2014+TESS, TESS+2015}.
Observational gaps, resulting from J0727 falling outside the field of view of TESS, segmented the available data into five distinct observation series (Series I-V, see Fig~\ref{fig:archival_lightcurve}).
The coverage, cadence, and data source for each series are detailed in Table~\ref{tab:TESS coverage}.

Series I and II employ light curves from the TESS-Gaia Light Curve \citep[TGLC,][]{Han+Brandt+2023+TGLC}.  To mitigate contamination from nearby stars, TGLC models Full Frame Images (FFIs) with effective point spread functions (PSFs) by leveraging astrometry and photometry from the Gaia DR3.
We extracted these light curves from FFIs using the {\tt tglc} package \citep{Han+Brandt+2023+TGLC}.
Observation of Sector 9 were excluded because J0727 was exactly positioned at the image periphery.

Series III and IV consist of light curves processed by the TESS-SPOC pipeline \citep{Caldwell+etal+2020+TESS-SPOC}, which extends the Science Processing Operations Center (SPOC) pipeline \citep{Jenkins+etal+2016+SPOC} to generate light curves from FFIs for targets selected via the TESS Input Catalog \citep[TIC,][]{Stassun+etal+2019+TESS_TIC}. Series V comprises light curves generated directly by the SPOC pipeline \citep{Jenkins+etal+2016+SPOC} from two-minute cadence target data. We retrieved light curves for Series III-V using the {\tt lightkurve} package \citep{Lightkurve_Collaboration+etal+2018+lightkurve}.

Concerning data processed with SPOC pipeline, TESS light curves contain two types of flux measurements: Simple Aperture Photometry (SAP) and Pre-search Data Conditioning SAP (PDCSAP).
The SAP flux is calculated by summing together the pixel fluxes within the optimal aperture set by TESS pipeline.
However, SAP flux retains systematic deviations inherent to the mission. 
The PDCSAP flux has been calibrated by removing long-term trends using Co-trending Basis Vectors \citep[CBVs,][]{Smith+etal+2012+CBVs, Stumpe+etal+2014+CBVs}.
While PDCSAP flux typically exhibits fewer systematic trends than SAP flux, this correction process also attenuate or remove genuine long-timescale astrophysical signals.
Consequently, PDCSAP flux is generally employed for analyzing short-timescale phenomena \citep[e.g. stellar flares,][]{Gunther+etal+2020+PDCSAP_flares}, whereas SAP flux is preferred for studying long-timescale variability \citep[e.g. outbursts,][]{Ilkiewicz+etal+2024+DN_outburst}. Both flux types were utilized in this work.

\begin{table}
    \centering
    \caption{Basic information of TESS observations for J0727.}
    \begin{tabular}{ccccc}
        \hline
        \hline
        Series & Sectors & Coverage(BJD-2457000) & Cadence(s) & Pipeline\\
        \hline
        I   & 1--8 & 1325.32--1541.99 & 1800 & TGLC\\
        II  & 10--13 & 1573.05--1682.34 & 1800 & TGLC\\
        III & 27--38 & 2036.29--2360.54 & 600  & TESS-SPOC\\
        IV  & 61--69 & 2964.02--3206.15 & 200  & TESS-SPOC\\
        V   & 87--90 & 3665.88--3773.67 & 120  & SPOC\\
        \hline
        \hline
    \end{tabular}
    \label{tab:TESS coverage}
\end{table}

\subsection{Gaia color and absolute magnitude} \label{subsec:Gaia CMD}
To characterize J0727 within the color-magnitude diagram (CMD), we derived its absolute magnitude and color using {\it Gaia} DR3 database \citep{Gaia_Collaboration+2016+performance,Gaia+DR3+2022}.
Owing to high signal-to-noise ratio (SNR) of its parallax, $\varpi=1.040\pm0.022$~mas, the distance to J0727 is well constrained at $d=962\pm20$~pc.
We estimated the extinction parameters for J0727 by utilizing the derived extinction and color-excess values of its neighboring sources, obtained from the {\it Gaia} Astrophysical Parameters Inference System \citep[Apsis,][]{GaiaDR3+Apsis+parameters}.
This approach follows similar methodologies applied to other Galactic variable stars \citep{Caiazzo+etal+2021+extinction,Lin+etal+2023+BLAP_NA}. 
We extracted a subset comprising the 100 nearest {\it Gaia} stars within a 0.5$\degree$ radius centered on the target. 
The statistical analysis of their Apsis extinction parameters yielded $A_{\rm G}=0.332\pm0.164$~mag and color excess $E(B_{\rm p}-R_{\rm p})=0.181\pm0.091$~mag.
Given apparent magnitude $G_{\rm p} = 14.658\pm0.007$~mag and color $B_{\rm p}-R_{\rm p}=0.200\pm0.024$~mag,
the dereddened color of J0727 is $(B_{\rm p}-R_{\rm p})_0=0.020\pm0.094$\,mag and its absolute magnitude is $M_{\rm G}=4.410\pm0.170 $\,mag.
These values place J0727 within the region occupied by nova-like stars on the {\it Gaia} CMD \citep{Abril+etal+2020+Gaia_CV_distribution}.

\section{Spectroscopic observations} \label{sec:spectroscopy}
\setcounter{footnote}{0}
A spectrum of target J0727 with resolution $R=3000$ was obtained on 18 May 2023 using Wide Field Spectrograph \citep[WiFeS,][]{Dopita+etal+2010+WiFeS}, mounted on the Australian National University (ANU) 2.3m Telescope at Siding Spring Observatory (Proposal ID: 2360140). 
The WiFeS data were reduced via {\tt FIGARO} \citep{Rinaldi+DelPozzo+2024+FIGARO} following standard procedures\footnote{\url{https://starlink.eao.hawaii.edu/docs/sun86.pdf}}. 
A sequence of FIGARO commands was applied to each frame for bias subtraction, cosmic-ray removal, flat-field correction, and extraction of individual 2D spectral slices.
Barrel distortion, predominantly along the vertical axis (i.e. spatial direction), was rectified using transformations derived from NeAr arc exposures.
The individual 2D slices were then aligned and underwent preliminary background subtraction.
To compensate for optical distortion and atmospheric dispersion, the slices were spatially remapped to ensure stellar centroids aligned along a common row.
The rows containing the stellar signal were then selected and summed to form the raw stellar spectrum. An equivalent number of adjacent background rows were similarly summed and subtracted from the stellar spectrum.

Each resultant stellar spectra derived from individual slices was wavelength-calibrated via 5th-order polynomial fits to the NeAr arc and then co-added to produce the total raw spectrum.
Flux calibration followed a sequence: the combined spectrum was first corrected for mean atmospheric extinction, then divided by a normalized featureless standard star template to balance instrumental response, and finally calibrated against flux standards from the CALSPEC spectra\footnote{\url{https://archive.stsci.edu/hlsps/reference-atlases/cdbs/current_calspec/}}.

\begin{figure*}
\centering
    \includegraphics[width=0.95\textwidth]{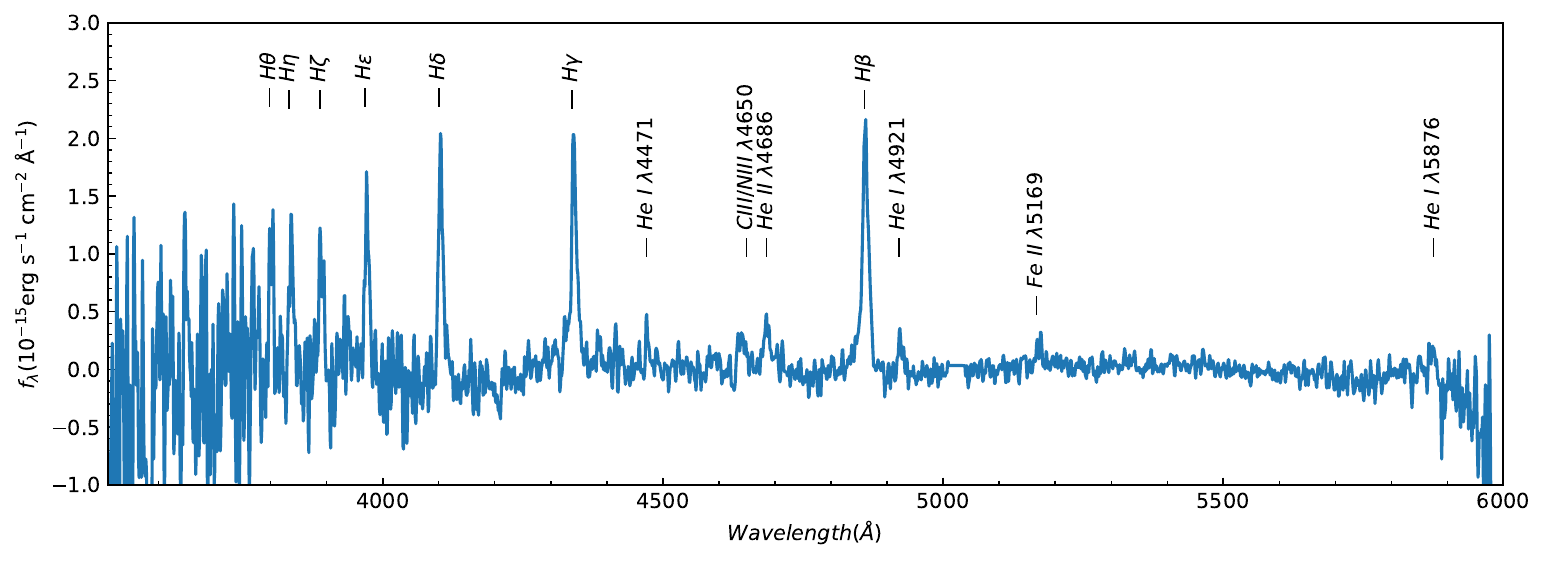}
    \caption{Continuum-subtracted spectrum of J0727. 
    The spectrum was obtained on the high-luminosity state of J0727.
    The strong Balmer lines, He I/II lines and metal lines (CIII/NIII $\lambda4650$ and Fe II $\lambda5169$) were marked.
    }
    \label{fig:spectrum}
\end{figure*}

A power-law function was fitted exclusively to the continuum regions of the spectrum, and a continuum-subtracted spectrum was then obtained by subtracting the best-fitting model from the spectrum (see Fig.~\ref{fig:spectrum}).
The high-state spectrum displays strong H Balmer lines, fainter He~I/II lines, and metal features including the C~III/N~III blend at $\lambda4650$ and Fe II $\lambda5169$.

\section{Computation of WWZ and Lomb-Scargle periodogram} \label{sec:computation of WWZ and LSP}

\begin{figure*}
\centering
    \includegraphics[width=0.95\textwidth]{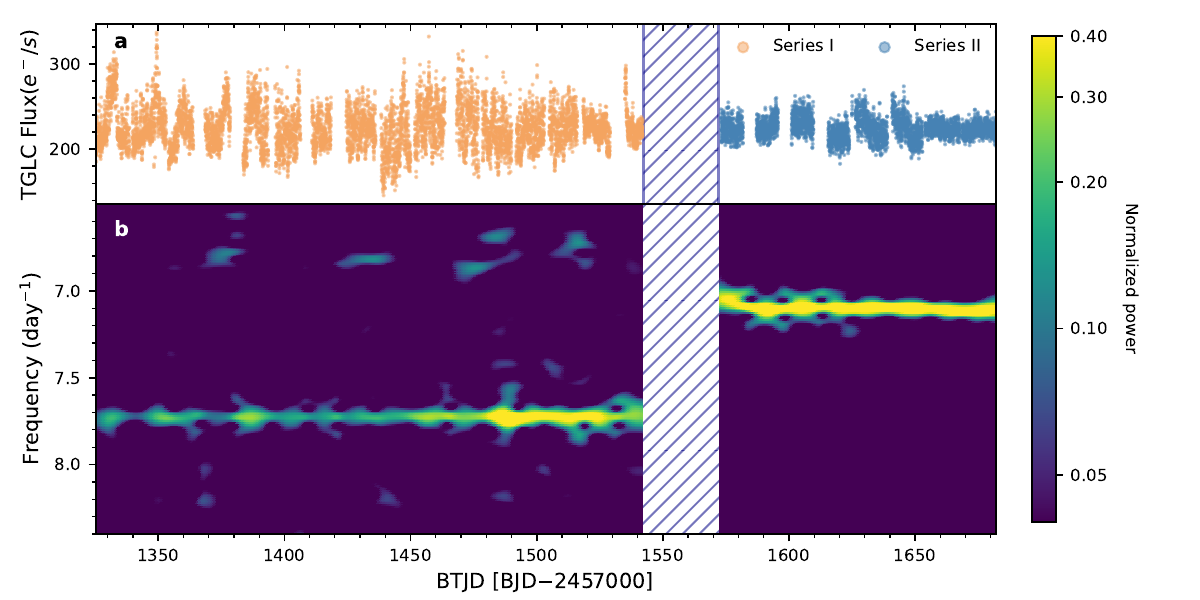}
    \caption{Normalized WWZ of J0727 during observation Series I and II.
    a, TGLC PSF light curves encompassing the onset of pSH. The photometric data points are from TESS observation Series~I (orange) and II (blue).
    b, Normalized WWZ diagram. Frequency and time bin widths are $0.01~{\rm day}^{-1}$ and 1.0~day, respectively.
    The blue shaded region denotes a significant observation gap resulting from J0727's position at the image periphery (See Appendix.~\ref{subsec:TESS photometry}).
    }
    \label{fig:wwz_I+II}
\end{figure*}

The WWZ analysis was performed over a frequency range of $6.4-8.4\,{\rm cycle\,day}^{-1}$ with a frequency bin width of 0.01\,${\rm day}^{-1}$.
We adopted a temporal step of $d\tau=1.0$\,day and a decay constant $c=1.2\times10^{-5}$ to optimally balance time and frequency resolution.
The WWZ diagrams for Series III and Series I-II are presented in Fig.~\ref{fig:wwz} and Fig.~\ref{fig:wwz_I+II}, respectively.
As shown in Fig.~\ref{fig:wwz_I+II}, the onset of positive superhump (pSH) fell into the observational gap between MJD~58541.99 to 58573.05.
Note that, although ASASSN observations span a wide temporal range (MJD 56776.02 to 60733.02), the WWZ analysis with these data is severely affected by daily aliasing and fails to resolve the relevant temporal features.

Regarding computation of LSPs, the Series III data (spanning 12 sectors) were divided into 24 segments according to the 13.7-day orbital gap. 
For each segment, we computed the LSP with a Nyquist frequency limit $\nu_{\rm Ny}=72\,{\rm day^{-1}}$ and an oversampling factor of $n_o=10$ .
Given the mean segment duration $T_{\rm seg} = 12.09$~day, the frequency step is set to $\delta \nu=1/(n_0T_{\rm seg})=0.008\,{\rm day^{-1}}$ \citep{VanderPlas+2018+understanding_lomb_scargle}.
Characteristic frequencies and their 1$\sigma$ uncertainties were statistically derived through 10,000 bootstrap iterations within adaptive frequency windows: $\nu_{\rm pSH}$ in 6.8--7.2\,${\rm day}^{-1}$, $\nu_{\rm orb}$ in 7.6--7.8\,${\rm day}^{-1}$ and $\nu_{\rm nSH}$ in 7.8--8.2\,${\rm day}^{-1}$. 
Consequently, this approach yielded orbital frequency $\nu_{\rm orb}=7.726\pm0.004~{\rm day}^{-1}$ and pSH frequency $\nu_{\rm pSH}=7.048\pm0.005~{\rm day}^{-1}$. 
In order to resolve the rapid evolution of nSH, $\nu_{\rm nSH}$ was determined by per individual segment rather than by staking the periodograms from all segments (see Fig.~\ref{fig:eruption_radius_decaying}).
However, given the observed frequency evolution trend, we omitted the nSH signal in the final segment, which was likely indistinguishable from the orbital frequency due to the coarse frequency resolution ($\Delta \nu \approx 1/T_{\rm seg}=0.084\,{\rm day}^{-1}$ here).

\section{Light-curve fitting} \label{sec:lightcurve fitting}

\begin{table*}
    \centering
    \caption{Best-fitting parameters for eruption E202104.}
    \begin{tabular}{lcc}
    \hline
    Parameter & Unit & Value \\
    \hline
    Peak flux, A & mJy & $3.666\pm0.043$ \\
    Time of the peak flux, $t_{\rm max}$ & BTJD & $2316.62\pm0.05$ \\
    Rising timescale, $\sigma_r$ & day & $10.63\pm0.75$ \\
    Decaying timescale, $\sigma_d$ & day & $16.71\pm0.75$ \\
    FWHM & day & 19.0 \\
    Duration ($3\sigma$- or $5\sigma$-cut) & day  & $44.99\pm2.48$ or $72.57\pm2.46$ \\
    Isotropic $g$-band energy ($3\sigma$- or $5\sigma$-cut) & $10^{39}$~erg & $4.76\pm0.21$ or $3.99\pm0.21$ \\
    Peak luminosity & $10^{33}$~erg/s & $3.82\pm0.16$ \\
    \hline
    \end{tabular}
    \label{tab:fitting params}
\end{table*}

\begin{figure*}
\centering
    \includegraphics[width=0.95\textwidth]{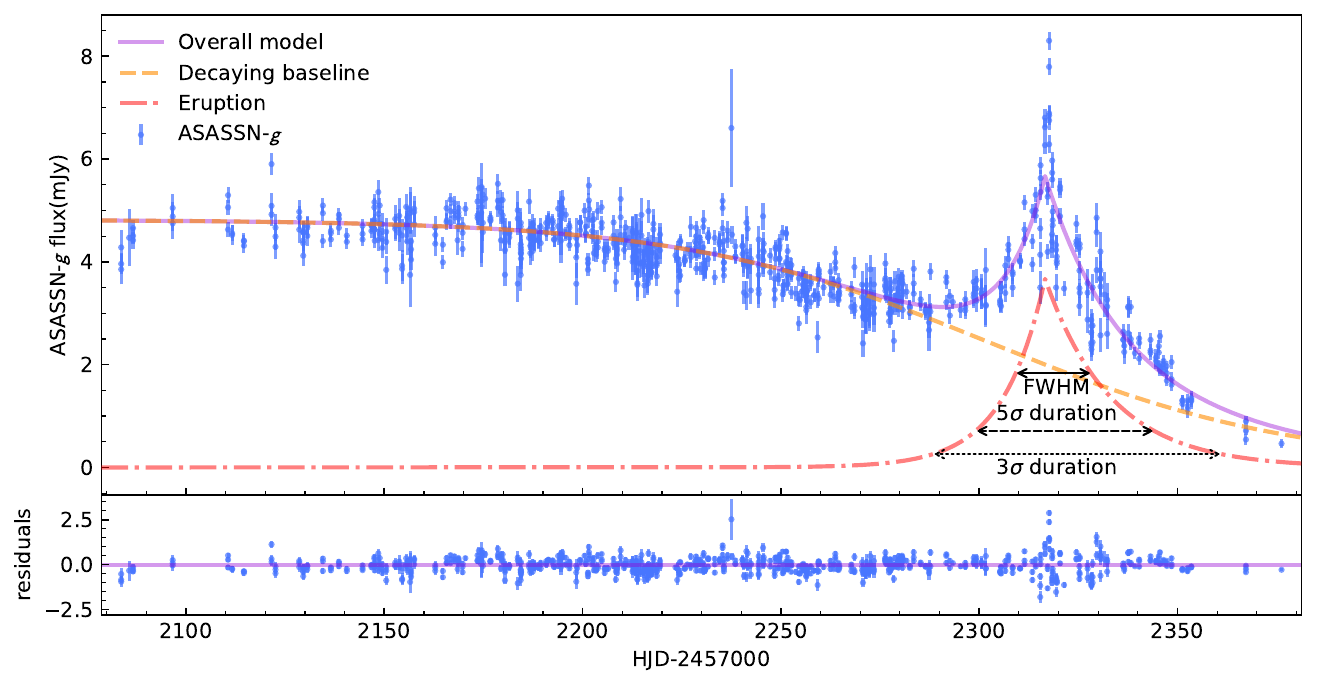}
    \caption{Light curve and best-fitting model for eruptive event E202104.
    a, ASASSN g-band light curve and best-fitting model. 
    The best-fitting model (purple solid line) consists of a double-exponential profile (for the eruption; red dot-dashed line) overlaid on a logistic function (for the decaying baseline; the orange dotted line).
    We marked the full width at half maximum (FWHM) of the eruption model (solid, double-sided arrow); the eruption duration defined by a 5$\sigma$ threshold above the baseline-subtracted quiescent level (dashed, double-sided arrow); and the duration defined by a 3$\sigma$ threshold (dotted, double-sided arrow).
    b,  flux residuals.
    }
    \label{fig:fitting lc}
\end{figure*}

To characterize the duration and energy release of the eruption concurrent with the brightness drop (BD2021), 
the contribution of eruption E202104 must be isolated by subtracting the underlying baseline of BD2021.
Empirically, the characteristic light curves of VY Scl-type brightness-drop events manifest a high-state plateau,  a low-state dip, and a smooth, sigmoidal transition connecting them.
This temporal profile can be mathematically described by a logistic function \citep{Verhulst+1838+logistic_function, Verhulst+1845+logistic_function},
\begin{equation*}
    F_{\rm base}(t)=F_{\rm high}-\frac{F_{\rm high}-F_{\rm low}}{1+e^{-k(t-t_0)}}~,
\end{equation*}
where $F_{\rm high}$ and $F_{\rm low}$ are the flux of high- and low- state, respectively; $t_0$ is the midpoint time of the brightness drop; and $k$ parametrizes the steepness of the decaying curve.

In order to fit the temporal profile of eruption E202104, we first consider the Linear Rise Exponential Decay \citep[LRED,][]{Hawley+Pettersen+1991+LRED_Mdwarf,Davenport+etal+2014+LRED_kepler} model, which is commonly employed in M-dwarf flare studies. However, this yields a positively skewed, asymmetric profile that does not match the near-symmetric characteristic of E202104.
On the other hand, conventional symmetric functions (e.g., Gaussian, Lorentzian) are incapable for capturing the eruption's sharp peak.
Instead, we utilized a double-exponential function adapted from gamma-ray burst modeling \citep{Norris+etal+1996+double_exponential_GRB}:

\begin{equation*}
    F_{\rm erup}(t) = \begin{cases}
        A \exp{[-(t_{\rm max}-t)/\sigma_r]}, \ \ t<t_{\rm max}, \\
        A \exp{[-(t-t_{\rm max})/\sigma_d]}, \ \ t>t_{\rm max},
    \end{cases}
\end{equation*}

where 
$A$ denotes the peak flux, and $t_{\rm max}$ is the time corresponding to the peak; 
$\sigma_r$ and $\sigma_d$ represent the rise ($t<t_{\rm max}$) and decay ($t>t_{\rm max}$) timescales, respectively.
This piecewise exponential parameterization effectively captures the sharp-peak feature while allowing for near-symmetric profiles with potentially distinct rise and decay timescales.
By constructing a compound model combining a logistic function and a double-exponential function, we fitted the ASASSN g-band light curve of J0727 (Fig.~\ref{fig:fitting lc}) using {\tt pymc} \citep{pymc+2023} and listed best-fitting parameters in Table~\ref{tab:fitting params}

\section{Energy release and duration of various eruptive events} \label{sec:eruption statistics}

\begin{figure*}
\centering
    \includegraphics[width=0.95\textwidth]{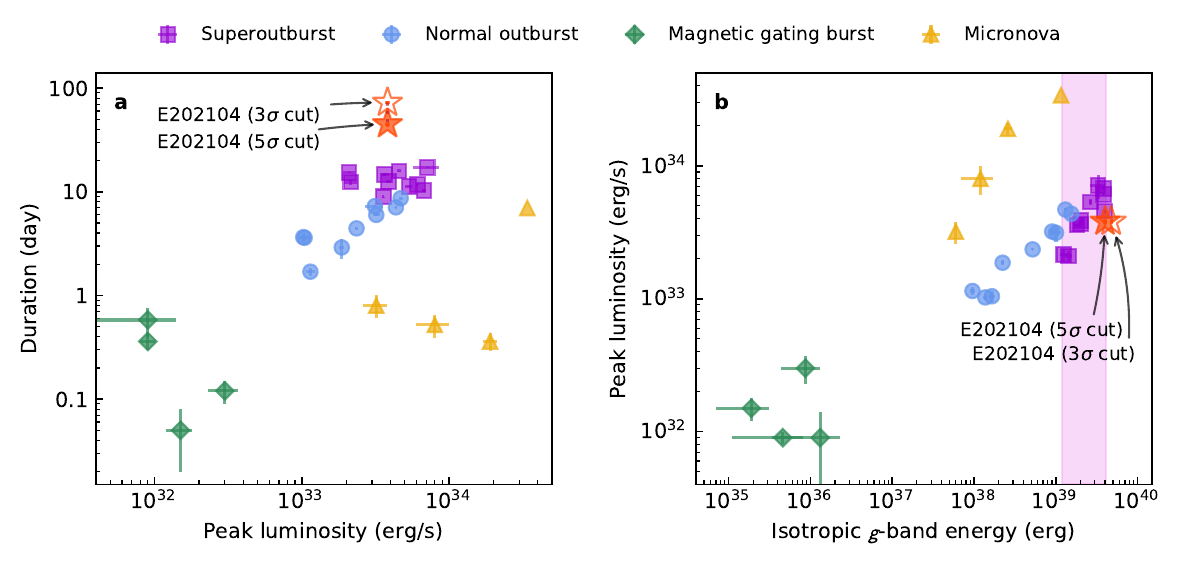}
    \caption{Duration-peak luminosity and peak luminosity-energy distribution for various types of eruptive events.
    a, Duration versus peak luminosity plane. 
    b, Peak luminosity versus energy plane.
    Symbols as in Fig.~\ref{fig:duration_energy_distribution}.
    }
    \label{fig:extended_eruption_distribution}
\end{figure*}

Aiming at investigating the physical origin of E202104, we compared its energy release, duration, and peak luminosity against various eruption events across different types of eruptive events, including magnetic gating bursts, micronovae, normal outbursts (NOs), and superoutbursts (SOs).
Observational properties for magnetic gating bursts  \citep{Ilkiewicz+etal+2024+DN_outburst,Scaringi+etal+2017+Nature+type2,Scaringi+etal+2022+Nature+type1,Littlefield+etal+2022+type2} and micronovae \citep{Scaringi+etal+2022+switch_type2,Veresvarska+etal+2024+bursting_type1} were obtained from literature.
To ensure consistent comparisons, we applied a similar measurement method to NOs and SOs in DN using TESS and ASASSN $g$-band observation data.

Here, our measurements of NOs and SOs rely on contemporaneous observational coverage from TESS and ASASSN-$g$.
TESS provides high-cadence photometry, while the ASASSN-$g$ light curves are used to calibrate the TESS photometry \citep{Scaringi+etal+2017+Nature+type2,Scaringi+etal+2022+switch_type2,Scaringi+etal+2022+Nature+type1,Ilkiewicz+etal+2024+DN_outburst}.
Through cross-match of DN catalog \citep{Otulakowska-Hypka+etal+2016+DN_statistic} with both the TESS and ASASSN-$g$ datasets, we selected outburst samples with well observation coverage that were not interrupted by gaps between TESS sectors.
Given that the superoutburst in WZ Sge-type stars typically exhibit higher energy than those in SU UMa-type stars \citep{Warner+1995+CV_book} and are potentially comparable to the event E202104,
we queried the Variable Star Index \citep[VSX,][]{Watson+Henden+Price+2006+VSX} in addition and manually added the target V0748 Hya \citep[identified as a WZ Sge-type star,][]{Uemura+etal+2010+V0748Hya} to the sample. The final outburst sample comprises 9 NOs and 10 SOs.

For each TESS sector, we selected contemporaneous photometric points between ASASSN-$g$ and TESS that were taken within 0.002~day of each other. 
Then a linear correlation was established between the TESS SAP flux (in $e^{-}/s$) and the corresponding ASASSN $g$-band flux density (in mJy). 
To ensure unit consistency with the measurements of magnetic gating bursts and micronovae, these flux densities in frequency space ($f_{\rm g,\nu}$) were converted to flux densities in wavelength space ($f_{\rm g,\lambda}$) using the transformation 
\begin{equation*}
f_{\rm g,\lambda}=f_{\rm g,\nu}|\frac{\partial\nu}{\partial\lambda}|=f_{\rm g,\nu}\times\frac{c}{\lambda_{\rm eff,g}^2}~,
\end{equation*}
where $c$ is the speed of light and $\lambda_{\rm eff,g}=4639.27\,\text{\AA}$ is the effective wavelength of the ASASSN $g$-band filter \citep{Rodrigo+Solano+Bayo+2012+SVO1.0,Rodrigo+etal+2024+SVO}.

The isotropic monochromatic $g$-band luminosity $L_{\rm g}$ was calculated as $L_{\rm g}=4\pi d^2 f_{\rm g,\lambda}\lambda_{\rm eff,g}$\citep{Scaringi+etal+2017+Nature+type2,Scaringi+etal+2022+switch_type2}, where $d$ is the source distance derived from Gaia DR3 parallax measurement \citep{Gaia+DR3+2022}.
While luminosity calculations for CVs typically incorporate a correction factor for disk inclination due to the radiative anisotropy of their accretion disks
\citep{Warner+1995+CV_book,Mayo+Wickramasinghe+Whelan+1980+disk_inclination,Paczynski+Schwarzenberg-Czerny+1980+inclination_disks}, we assume the isotropic luminosity here to enable direct comparison with magnetic gating bursts, micronovae, and the peculiar eruptive event E202104.

By interpolating the calibrated TESS light curves,
each outburst event (NO \& SO) was isolated by applying a threshold set at 5$\sigma$ above the quiescent flux level, enabling preliminary determination of outburst duration.
Here, $\sigma$ represents the standard deviation of quiescence flux.
Peak luminosity was defined as the maximum luminosity during the outburst, and the isotropic $g$-band energy was calculated as the temporal integral of the luminosity over the determined outburst duration. 
To estimate the uncertainties in the isotropic monochromatic energy, peak luminosity, and duration, we performed 10,000 bootstrap iterations, incorporating 1$\sigma$ Gaussian errors in both the distance estimates and the photometry for all epochs.
This pipeline was also applied to a subset of micronovae and magnetic gating bursts, and yielded luminosities consistent with published results \citep{Scaringi+etal+2022+Nature+type1,Scaringi+etal+2022+switch_type2}, thus validating the methodology.

For the eruption E202104,
besides the standard 5$\sigma$ threshold,
an additional 3$\sigma$ cut were applied to determine its duration and isotropic energy,
aiming at offsetting its slowly developed temporal profile.
As shown in Figures \ref{fig:duration_energy_distribution}, the exceptionally long duration of E202104 distinguishes it from other types of eruptive events. 
However, in the peak luminosity versus isotropic energy plane (right panel of Fig. \ref{fig:extended_eruption_distribution}), 
E202104 occupies a region overlapping with the most energetic SOs produced by DNe.
This means that the peak luminosity and energy release of E202104 are comparable to those of SOs, which are believed to consume most of the mass stored in an accretion disk.

\bibliography{reference}
\bibliographystyle{aasjournalv7}

\end{document}